\documentstyle[12pt]{article}
\def \ms {{\overline{\mbox{MS}}}}

\newcommand{\prepr}[1]
{\begin{flushright} {\bf #1} \end{flushright} \vskip 1.5cm}
\newcommand{\titul}[1] {\begin{center}{\large\bf #1 } \end{center}\vskip 1.cm}
\newcommand{\autor}[1] {\begin {center} {\large \lineskip .5em #1 }
                        \end   {center} }
\newcommand{\lugar}[1] {\begin{center} {\it #1} \end{center}}
\newcommand{\abstr}[1] {{\begin{center} \vskip .5cm {\bf Abstract
                        \vspace{0pt}} \end{center}}\begin{quote} #1
                        \end{quote}}
\newcommand{\z}{&&\hspace*{-1cm}}
\begin{document}

\begin{titlepage}
\prepr{US-FT/41-96\\ September 1996}
\titul{The DIS cross-sections
ratio $R= \sigma_L/\sigma_T$
at small x \\
from HERA data.
}

\autor{A.V. Kotikov $^a$}
\lugar{Laboratoire de Physique Theorique ENSLAPP\\ LAPP, B.P. 100,
F-74941, Annecy-le-Vieux Cedex, France}
\autor{G. Parente $^b$}
\lugar{Departamento de F\'\i sica de Part\'\i culas\\
Universidade de Santiago de Compostela\\
15706 Santiago de Compostela, Spain}
\abstr{We extract the deep inelastic scattering cross-sections
ratio $R= \sigma _L/\sigma _T$
in the range $10^{-4} \leq x \leq  10^{-1}$
from $F_2$ HERA data using very simple relations based on perturbative QCD.  
The result depends on only one parameter
$\delta$, being $x^{-\delta}$ the behavior of the parton
densities at low $x$, which has been determined recently with
a good accuracy by the H1 group. 

\vskip 0.5 cm

}
\end{titlepage}
\newpage
In recent years the behaviour of deep inelastic
lepton-hadron scattering (DIS) 
at the small values of Bjorken
variable $x$, has been intensively studied.
The present letter is devoted to the behaviour of 
the ratio of cross-sections of the absorption of a longitudinal- and
transverse-polarized photon by hadron: $R= \sigma _L/\sigma _T$,
 at small values of $x$.
The ratio $R$, which may be represented as the combination of the
longitudinal $F_L(x,Q^2)$ and transverse $F_2(x,Q^2)$ DIS structure
functions (SF):
\begin{eqnarray}
R(x,Q^2)~=~ \frac{F_L(x,Q^2)}{F_2(x,Q^2) ~-~ F_L(x,Q^2)} 
\label{0.1} 
\end{eqnarray}
is a very sensitive QCD characteristic because it is
equal to zero in the parton model with spin$-1/2$ partons
(it is very large with spin$-0$ partons). 
However, modern DIS experimental data (see  the review in \cite{1}) are not
accurate enough to determine $R(x,Q^2)$. 
In addition, at
small values of $x$, $R$ data are not yet 
available,
as they require 
a rather cumbersome procedure (see \cite{1.5}, for example)
for the extraction from the experiment.

We study the behaviour of
$R(x,Q^2)$ at
small values of $x$, using the H1
data \cite{F2H1, H1PREP95} and the method 
\cite{2}
of replacement of the Mellin convolution by ordinary products.
By analogy with the case of the gluon distribution function (see 
\cite{H1PREP95, ZEUSGLU95, EKL, KOPA}) 
it is possible to obtain the relation between $F_L(x,Q^2)$, $F_2(x,Q^2)$
and $dF_2(x,Q^2)/dlnQ^2$ at small $x$. Thus, the small $x$ behaviour of
the ratio $R(x,Q^2)$ can be extracted directly from the measured values
of $F_2(x,Q^2)$ and its derivative.
These extracted values of $R$ may be well considered as
{\it new
small $x$ ``experimental data''} $^c$. Moreover,
when experimental data for $R$ at small $x$ become
available with a good accuracy, a violation of this {\it exactly perturbative}
relation will be an indication of the
importance of other effects as higher twist contribution and/or
of non-perturbative QCD dynamics at small $x$.

We follow the notation of our previous work in refs. \cite{KOPA,KOPAFL}. 
The singlet quark
$s(x,Q^2_0)$ and gluon $g(x,Q^2_0)$ parton distribution functions 
(PDF) $^d$
at some $Q^2_0$ are parameterized by
  (see, for example, \cite{4}):
 \begin{eqnarray} 
p(x,Q^2_0) & = & A_p
x^{-\delta} (1-x)^{\nu_p} (1+\epsilon_p \sqrt{x} + \gamma_p x)
~~~~(\mbox{hereafter } p=s,g).
\label{1} \end{eqnarray}

The value of $\delta$ is a matter of controversy.
  The ``conventional'' choice is $\delta =0$, which leads to a non-singular
behaviour of the PDF
when $x \rightarrow 0$. Another value,
$ \delta  \sim \frac{1}{2}$, was obtained in the studies performed in
ref. \cite{6} as
the sum of the leading powers of $\ln(1/x)$ in all orders of perturbation
theory. 
Experimentally, recent NMC data
\cite{7} favor small values of $\delta$. This result is also in
agreement with
present data for $pp$ and $\overline p p$ total cross-sections 
(see \cite{8}) and
corresponds to the model of Landshoff and Nachtmann pomeron
 \cite{9} with the exchange of a pair of non-perturbative gluons,
yielding  $\delta =0.086$.  However, the new HERA data
\cite{F2H1, ZEUSGLU95}  
 prefer $ \delta \geq 0.2$.
For example, the $\delta$ value obtained recently by H1 group
\cite{F2H1} seems to depend slowly on $Q^2$ values. Its average value
increases from $\delta  = 0.228$ at $Q^2 = 8.5$ GeV$^2$ to 
$\delta  = 0.503$ at $Q^2 = 800$ GeV$^2$.

Further, we restrict the analysis to the case of large $\delta $ values
(i.e. $x^{-\delta} \gg 1$ ) following recent H1 data \cite{F2H1}.
The more complete analysis concerning to the
extraction of the longitudinal SF $F_L(x,Q^2)$, may be found in
\cite{KOPAFL}, where we took into account also the case $\delta \sim
0$ corresponding to the standard pomeron.\\ 

 Assuming the {\it Regge-like behaviour} for the gluon distribution and
 $F_2(x,Q^2)$ at $x^{-\delta} \gg 1$:
$$g(x,Q^2)  =  x^{-\delta} \tilde g(x,Q^2),~~ 
F_2(x,Q^2)  =  x^{-\delta} \tilde s(x,Q^2), $$
 we obtain the
following equation for the $Q^2$ derivative of
the SF $F_2$ $^e$:
 \begin{eqnarray} \z \frac{dF_2(x,Q^2)}{dlnQ^2}  =
-\frac{1}{2} 
x^{-\delta} \sum_{p=s,g}
\Bigl( 
 r^{1+\delta}_{sp}(\alpha) ~\tilde p(0,Q^2) +
 r^{\delta}_{sp}(\alpha)~ x \tilde p'(0,Q^2)  +
 O(x^{2}) \Bigr), \nonumber \\ \z
F_L(x,Q^2)  =
x^{-\delta} \sum_{p=s,g}
\Bigl( 
 r^{1+\delta}_{Lp}(\alpha) ~\tilde p(0,Q^2) +
 r^{\delta}_{Lp}(\alpha)~ x \tilde p'(0,Q^2)  +
 O(x^{2}) \Bigr) , 
\label{2.1} \end{eqnarray}
 where $r^{\eta}_{sp}(\alpha) $ and $r^{\eta}_{Lp}(\alpha) $
 are the combinations
of the anomalous dimensions of Wilson operators 
$\gamma^{\eta}_{sp}= \alpha \gamma^{(0),\eta}_{sp} + \alpha ^2 
\gamma^{(1),\eta}_{sp} + O(\alpha ^3)$
and Wilson 
coefficients $^f$
$ \alpha B_L^{p,\eta} 
\Bigl(1+ \alpha R_L^{p,\eta} \Bigr)  + O(\alpha ^3)$
and $ \alpha B_2^{p,\eta}  + O(\alpha ^2)$
of the $\eta$
"moment"  (i.e., the corresponding variables extended
from integer values of argument to non-integer ones):
 \begin{eqnarray} \z
r^{ \eta}_{Ls}(\alpha) ~=~ \alpha 
  B^{s,\eta}_{L} \biggl[1 + \alpha \biggl(
  R^{s,\eta}_{L} - B_2^{s,\eta}   \biggr) 
  \biggr] + O(\alpha ^3),  \nonumber  \\  \z
r^{\eta}_{Lg}(\alpha) ~=~ \frac{e}{f}  \alpha 
  B^{g,\eta}_{L} \biggl[1 + \alpha \biggl(
  R^{g,\eta}_{L} - B_2^{g,\eta} B^{s,\eta}_{L} / B^{g,\eta}_{L}   \biggr) 
  \biggr]   + O(\alpha ^3), \nonumber \\ \z
r^{ \eta}_{ss}(\alpha) ~=~ \alpha 
  \gamma^{(0),\eta}_{ss} + \alpha^2 \biggl(
  \gamma^{(1),\eta}_{ss} + B_2^{g,\eta} \gamma^{(0),\eta}_{gs} +
  2\beta_0 B_2^{s,\eta}
  \biggr) + O(\alpha ^3),  \label{2.2}  \\  \z
r^{\eta}_{sg}(\alpha) ~=~ \frac{e}{f} \biggl[ \alpha 
  \gamma^{(0),\eta}_{sg} + \alpha^2 \biggl(
  \gamma^{(1),\eta}_{sg} + B_2^{s,\eta} \gamma^{(0),\eta}_{sg}
                         + B_2^{g,\eta} \bigl(2\beta_0 + 
  \gamma^{(0),\eta}_{gg} - \gamma^{(0),\eta}_{ss} \bigr) \biggr)
  \biggr] + O(\alpha ^3), \nonumber  \end{eqnarray} 
 and
$$ \tilde p'(0,Q^2) \equiv \frac{d}{dx} \tilde p(x,Q^2) \mbox{ at }
x=0,$$
where $e = \sum_i^f e^2_i$ is
the sum of squares of 
quark charges.

With accuracy of $O(x^{2-\delta})$,  
we have for Eq.(\ref{2.1}) 
 \begin{eqnarray}    \z    \frac{dF_2(x,Q^2)}{dlnQ^2}  = -\frac{1}{2} \biggl[
r^{1+\delta}_{sg} {(\xi_{sg})}^{-\delta}
   g(x/\xi_{sg},Q^2) + r^{1+\delta}_{ss}
F_2(x,Q^2)  + (r^{\delta}_{ss} - r^{1+\delta}_{ss}) x^{1-\delta}
\tilde s'(x,Q^2)
 \biggr] \nonumber \\ \z
~~~~~~~ ~~~~~~~~
+~ O(x^{2-\delta}) , \label{5} \\
   \z \nonumber \\ \z   
F_L(x,Q^2)  = r^{1+\delta}_{Lg} {(\xi_{Lg})}^{-\delta}
   g(x/\xi_{Lg},Q^2) + r^{1+\delta}_{Ls}
F_2(x,Q^2)  + (r^{\delta}_{Ls} - r^{1+\delta}_{Ls}) x^{1-\delta}
\tilde s'(x,Q^2) \nonumber \\ \z
~~~~~~~ ~~~~~~~~
+~ O(x^{2-\delta}) , \label{5.1}
\end{eqnarray}
with $\xi_{sg} =  r^{1+\delta}_{sg} /r^{\delta}_{sg}$ and 
$\xi_{Lg} =  r^{1+\delta}_{Lg} /r^{\delta}_{Lg}$.

From Eq.(\ref{5}) and (\ref{5.1}) one can obtain $F_L$ as a function
of $F_2$ and the derivative

\begin{eqnarray} \z
F_L(x, Q^2)  = -
\xi ^{\delta}
  \biggl[ 2 \frac{ r^{1+\delta}_{Lg}}{ r^{1+\delta}_{sg}} 
\frac{d F_2(x \xi , Q^2)}{dlnQ^2}
+ \biggl( r^{1+\delta}_{Ls} - \frac{ r^{1+\delta}_{Lg}}{ r^{1+\delta}_{sg}}
r^{1+\delta}_{ss} \biggr) F_2(x \xi ,Q^2)  \nonumber \\ \z
 ~+~  
O(x^{2-\delta},\alpha x^{1-\delta}) \biggr] , 
\label{8}\end{eqnarray}
where the result is restricted to $O(x^{2-\delta}, \alpha
 x^{1-\delta})$. 
To arrive to the above equation we have 
performed the substitution
$$ \xi_{sg}/\xi_{Lg} \to  \xi =
\gamma_{sg}^{(0),1+\delta} B^{g,\delta}_L /\gamma_{sg}^{(0),\delta}
B^{g,1+\delta}_L  $$ 
and neglected the term $\sim \tilde s'(x \xi_{sg},Q^2)$.

This replacement is very useful. The NLO anomalous dimensions
$\gamma_{sp}^{(1),n}$ are 
singular
 in both points, $n=1$ and $n=0$, and their presence
into the arguments of $\tilde p(x,Q^2)$ makes the numerical
agreement between this approximate formula and the exact calculation worse
(we have
checked this point using some MRS sets \cite{4} of parton
distributions).

Using NLO approximation of $r^{1+\delta}_{sp}$ and $r^{1+\delta}_{Lp}$
for concrete values of $\delta = 0.5$ 
and $\delta = 0.3$ we obtain (for f=4 and $\overline{MS}$ scheme):
\begin{eqnarray} \z  
~\mbox{ if }~ \delta =0.5 \nonumber \\ \z
F_L(x, Q^2)  = \frac{0.87}{ 1 + 22.9 \alpha} \Biggl[
  \frac{d F_2(0.70 x , Q^2)}{dlnQ^2} + 
4.17 \alpha F_2(0.70 x,Q^2)  \Biggr]  +
O(\alpha^2,x^{2-\delta},\alpha x^{1-\delta}) ,
\label{10.1} 
\\ \z \nonumber 
\\ \z
~\mbox{ if }~ \delta =0.3 \nonumber \\ \z
F_L(x, Q^2)  = \frac{0.84}{1 + 59.3 \alpha } \Biggl[ 
 \frac{d F_2(0.48 x, Q^2)}{dlnQ^2} + 3.59 \alpha 
F_2(0.48 x , Q^2) \Biggr]
+ O(\alpha^2 \!\!, x^{2-\delta} \!\!,\alpha x^{1-\delta}). 
\!         
\label{10.3}
\end{eqnarray}\\

With the help of Eqs. (\ref{0.1}) and  (\ref{10.1})-(\ref{10.3})
we have extracted the ratio $R(x,Q^2)$
from H1 1994
data \cite{F2H1}, determining the slopes dF$_2$/dlnQ$^2$
from straight line fits as in ref. \cite{H1PREP95,ZEUSGLU95}.
In the present calculation only statistical errors have been taken into
account, and we have used $\Lambda^{(4)}_{\ms}=225 MeV$ in the calculation
of the running coupling constant $\alpha_s(Q^2)$ at two loops.

Figure 1a shows the extracted ratio $R$ at
$Q^2 = 20$ GeV$^2$ for two different values of the parameter $\delta$.
It also shows BCDMS \cite{BCDMS} and preliminary
CCFR (see \cite{CCFR}) data points where the errors are very much larger.
For comparison we have also plotted various predictions for $R$ using
QCD formulas at O($\alpha_s^2$) \cite{FLNLO} and parton densities extracted
from fits to HERA data. The large difference between the
result from MRS(G) and the latest set MRS(R1) \cite{MRS96} 
shows, as it is expected, the large effect on R of the
unknown of the gluon distribution at small $x$.

In figure 1a one can also see that the result from MRS(R1) fits very
well the points obtained with $\delta=0.5$ for the lowest $x$ data, although
it fails to account for the highest $x$ bins.
The calculation with MRS(D-) is also statistically compatible with our data.

By other part recent theoretical predictions
on $R$ based on conventional NLO DGLAP evolution analysis of HERA data
(LBY) \cite{LBY} and on the dipole picture of BFKL dynamics
(NPRW) \cite{NPRW}, both finding values $\delta=0.3$,
lie closer to the data points
obtained with $\delta=0.3$ Eq. (\ref{10.3}).

Finally Fig. 1b shows $R$ 
for $\delta=0.3$ (the value favoured by H1 data \cite{F2H1}) and
at three different $Q^2$ values in comparison with
the SLAC R(1990) parametrization \cite{RSLAC}. One can see
the very good agreement at $x \leq 10^{-2}$ even if only the
statistical errors are taken into account.

Notice that the points at the same $x$ and different $Q^2$
are correlated by the form in which the derivative
term $dF_2/dlnQ^2$ is determined.\\

In summary, we have presented  Eqs. (\ref{0.1}) and
(\ref{8})-(\ref{10.3}) for the
extraction of the ratio $R= \sigma _L/\sigma _T$ at small $x$ from the SF
$F_2$ and its
$Q^2$ derivative. These equations provide the possibility of the non-direct
determination of $R$. This is important since the direct
extraction of $R$ from experimental data is a cumbersome procedure
(see \cite{1.5}).
Moreover, the fulfillment of  Eqs. (\ref{0.1}), (\ref{8})-(\ref{10.3}) by
DIS experimental data is a cross-check of perturbative QCD at small values of
$x$. Our formulas can also be used as a parametrization of $R$
as a function of the most widely used phenomenological $F_2$.

We have found
that 
the results  depend 
on the concrete value
of the slope $\delta $.
In the case $\delta = 0.3$, which is very close to the values obtained
by H1 group \cite{F2H1} at the considered $Q^2$ interval, we found
very good agreement with the SLAC parametrization \cite{RSLAC} and
also a relatively good agreement with the studies
based on NLO DGLAP
and  BFKL dynamics (see \cite{LBY} and
\cite{NPRW}, respectively). However the calculation performed
with the latest sets of HERA parton densities using perturbative
QCD at second order (see MRS(R1) curve in Fig. 1a) predicts an
slightly higher value for R.

\vskip 0.5 cm
%
%
%
%

This work was supported in part by CICYT and by Xunta de Galicia.
We are grateful to J.W. Stirling for providing
the parton distributions used in this work, and to 
A. Bodek and M. Klein for discussions.

\vskip 0.5 cm

%
%
a) Electronic address: KOTIKOV@LAPPHP8.IN2P3.FR.
  On leave of absence from Particle Physics Laboratory, JINR, Dubna, Russia.

b) Electronic adress: GONZALO@GAES.USC.ES

c) Although with the theoretical prejudice contained in the
  starting QCD relation.

d) We use PDF multiplied by $x$
  and neglect the nonsinglet quark distribution at small $x$.

e) Hereafter we use $ \alpha(Q^2)= \alpha_s(Q^2)/{4 \pi}$.

f) Because we consider here $F_2(x,Q^2)$ but not the singlet quark
  distribution.

%
%
\noindent{\bf Figure captions }
\vspace{1.cm}
%


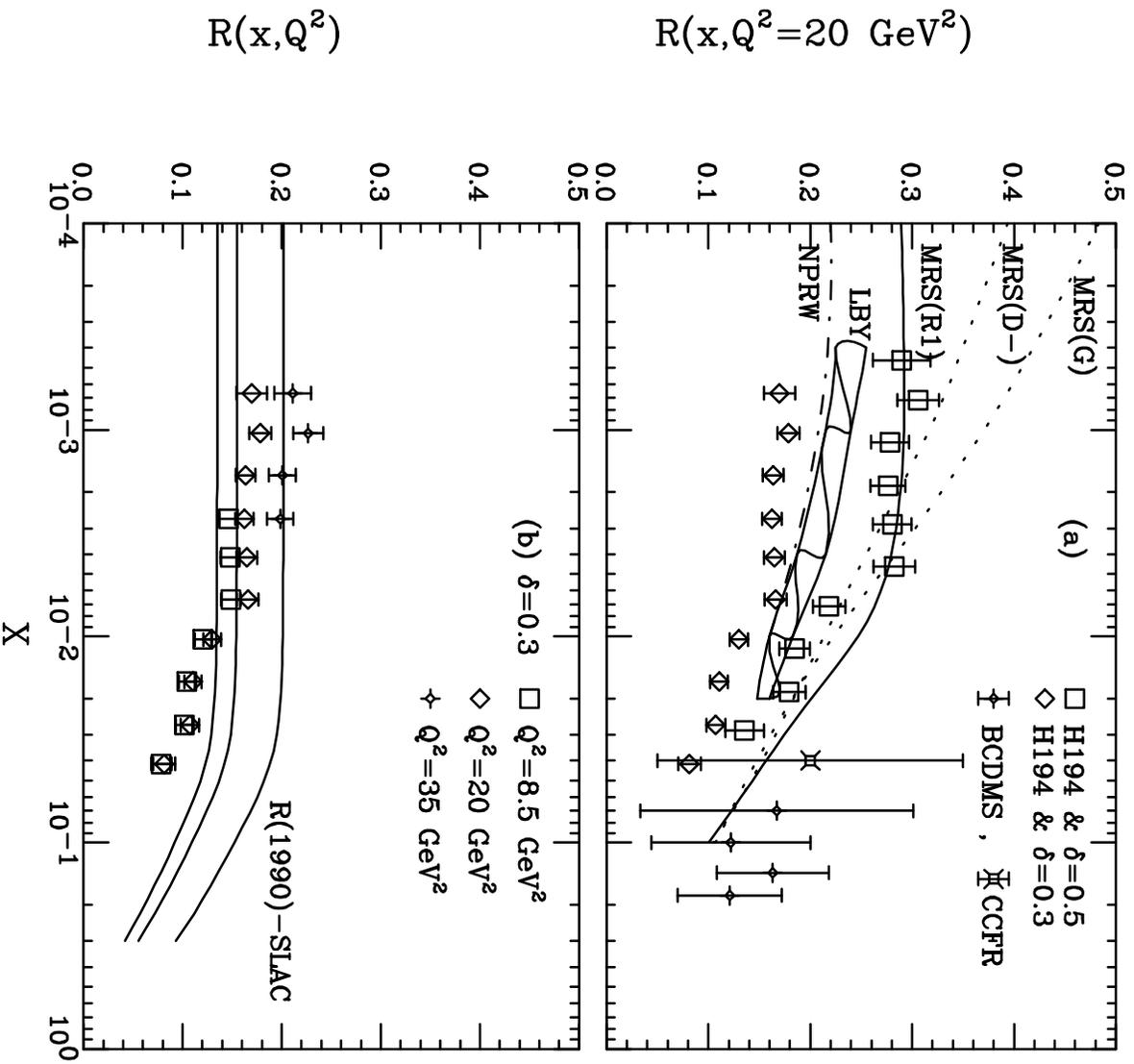
\noindent{Figure 1: The ratio $R= \sigma _L/\sigma _T$ at small $x$. 
The points were extracted
from Eqs. (\ref{0.1}), (\ref{10.1}) and (\ref{10.3}) 
using H1 \cite{F2H1, H1PREP95} 
data.
The dashed-dotted line (NPRW) is the prediction of Saclay group \cite{NPRW}
based on the dipole picture of BFKL dynamics. 
The band represent the uncertainty from the DGLAP analysis of HERA
data by \cite{LBY}.
It is also shown BCDMS data \cite{BCDMS} points at high $x$
 and the preliminary CCFR data point from \cite{CCFR}.
The solid lines in Fig. 1b are the SLAC R1990 parametrization
\cite{RSLAC} at $Q^2=8.5,20$ and $35$ GeV$^2$
(lower curve corresponds to lower $Q^2$ value).
}

\vspace{1.cm}
\end{document}